\documentstyle[11pt]{article}
\newfont{\Bbb}{msbm10 scaled\magstep1}
\newcommand{\e}{\mbox{e}}
\newcommand{\id}{\mbox{{\it {\bf 1}}}}
\let\be\beta

\def\beq{\begin{equation}}\def\eeq{\end{equation}}
\def\be{\begin{displaymath}}\def\ee{\end{displaymath}}

\newcommand{\RR}{\hbox{\Bbb R}}\newcommand{\SS}{\hbox{\Bbb S}}

\let\pa\partial

\newcommand{\Proof}{\noindent{\bf Proof}\quad}

\newtheorem{theorem}{Theorem}[section]
\newtheorem{prop}[theorem]{Proposition}

\newtheorem{defi}[theorem]{Definition}
\newcommand{\bea}{\begin{eqnarray}}
\newcommand{\eea}{\end{eqnarray}}
\begin{document}
%
%
\begin{center}
{\Large{\bf Hidden symmetry of the quantum Calogero-Moser system}}
\vspace{.5cm}

Vadim B. Kuznetsov\vspace{.3cm} \\
Institute of Mathematical Modelling$\,{}^{1,2,3}$\\
Technical University of Denmark,\\
DK-2800 Lyngby, Denmark
\end{center}
\footnotetext[1]{Supported by the grant from Forskerakademiet through
a Guest Professorship.}
\footnotetext[2]{From 1st of October 1995 the address is:
Centre de recherches math\'ematiques, Universit\'e de
Montr\'eal, C.P. 6128, succ. Centre-ville, Montr\'eal,
Qu\'ebec, H3C 3J7, Canada.}
\footnotetext[3]{On leave of absence from Department of Mathematical and
Computational Physics, Institute of Physics,
St.Petersburg University, St.Petersburg 198904, Russia.}
\vspace{.5cm}
\centerline{\bf Abstract}
\vspace{.2cm}
Hidden symmetry of the quantum Calogero-Moser system with the inverse-square
potential is explicitly demonstrated in algebraic sense. We find the
underlying algebra explaining the super-integrability phenomenon for this
system. Applications to related multi-variable Bessel functions
are also discussed.
\vskip 2cm
solv-int/9509001
\vskip 10cm
\pagebreak
%
%
\section{Introduction}
\noindent
It is well-known \cite{W} that the rational $N$-particle
Calogero-Moser problem in classical
mechanics is superintegrable. It means that to each of the integrals
of motion $H_k$, $k=1,\ldots,N$, in involution w.r.t. the standard Poisson
bracket one can find a set of $N-1$ additional algebraic, functionally
independent integrals $B^{(k)}_r$,
$r=1,\ldots,N-1$, which are all in involution with the $H_k$. So, generally,
one has a $N\times N$ table of the form:
$$
\left[\matrix{H_1    & B^{(1)}_1 &  \cdots & B^{(1)}_{N-1} \cr
              H_2    & B^{(2)}_1 &  \cdots & B^{(2)}_{N-1} \cr
              \vdots & \vdots    &  \ddots & \vdots         \cr
              H_N    & B^{(N)}_1 &  \cdots & B^{(N)}_{N-1}  }\right].
$$
The entries of the first column in this table are distinguished by the
fact that each of them defines a superintegrable system, i.e. the $k$th
entry $H_k$ is in involution with those in the first column and in the
$k$th row. Each entry of the rest of the table, so anyone of the
$B^{(k)}_r$'s, defines an integrable system with the integrals of motion
being some functions of the $H_k$'s.

In the present paper we show how to construct the same kind of
table for the {\it quantum}
Calogero-Moser system with the inverse square potential.
The `quantum table' will have the same properties as the
`classical table', i.e. the first column will give us the
partial differential operators (PDOs) of the $k$th order
($k=1,\ldots,N$) with the
superintegrability property, while $B^{(k)}_r$'s will be the
PDOs of the following orders: order$(B^{(k)}_r)=k+r-1$.
We derive the non-linear algebraic relations for all these
operators.
%
%
\section{Operators of Dunkl's type}
\setcounter{equation}{0}
\noindent
Let $P_{ij}$ be the permutation operators acting on the indices
$i$ and $j$, i.e.
the generators of the permutation group $\SS_N$ of $N$ numbers
($i,j=1,\ldots,N$):
\beq
P_{ij}=P_{ji}\,,\qquad (P_{ij})^2=\id\,,
\qquad
P_{ij}P_{jl}=P_{jl}P_{li}=P_{li}P_{ij}\,.
\label{2.1}
\eeq
Hereafter all indices run from $1$ to $N$, unless
otherwise stated.

Introduce the operators
\beq
\Delta_i=\sum_{j\neq i}\frac{1}{x_{ij}}(\id-P_{ij})\,,\qquad i=1,\ldots,N,
\label{2.2}\eeq
where we use the notation
\be
x_{ij}=x_i-x_j\,,
\ee
and have the following algebra for the $P_{ij}$ and $x_k$:
\bea [x_i,x_j]&=&0\,,\nonumber\\
P_{ij}x_k&=&x_kP_{ij}\qquad \mbox{if}\qquad \{k\}\cap\{i,j\}=\emptyset\,,
\label{2.3}\\
P_{ij}x_j&=&x_iP_{ij}\,.\nonumber\eea
Then it is straightforward to verify the following
relations for the operators $P_{ij}$, $\Delta_k$, and $x_l$.
\begin{prop}
\bea
(a)&& P_{ij}\Delta_k=\Delta_kP_{ij}\qquad\mbox{if}\qquad\{k\}\cap\{i,j\}
=\emptyset\,,\nonumber\\
&&\, P_{ij}\Delta_j=\Delta_iP_{ij}\,,\label{2.4}\\
(b)&& [\Delta_i\Delta_k,P_{ik}]=0\quad\Leftrightarrow\quad
[\Delta_i,\Delta_k]=0\,,\label{2.5}\\
(c)&& [x_i,\Delta_j]=(1-\delta_{ij})P_{ij}-\delta_{ij}\sum_{k\neq i}P_{ik}\,.
\label{2.6}\eea
\end{prop}
\vskip 1mm

\noindent
Introduce the differential operators:
\be
\pa_i=\frac{\pa}{\pa x_i}\,,\qquad [\pa_i,x_j]=\delta_{ij}\,,
\ee
which are closed together with $P_{ij}$ to the same kind of algebra as
in (\ref{2.3}):
\bea [\pa_i,\pa_j]&=&0\,,\nonumber\\
P_{ij}\pa_k&=&\pa_kP_{ij}\qquad \mbox{if}\qquad \{k\}\cap\{i,j\}=\emptyset\,,
\label{2.7}\\
P_{ij}\pa_j&=&\pa_iP_{ij}\,.\nonumber\eea
It is not difficult to verify the following algebraic relations
for the operators $\pa_i$, $\Delta_j$, and $P_{kl}$.
\begin{prop}
\bea
(a)&& [\pa_i,\Delta_j]=\frac{1}{x_{ij}^{\;2}}(\id-P_{ij})+\frac{1}{x_{ij}}
(\pa_i-\pa_j)P_{ij}\,,\quad i\neq j\,,\nonumber\\
(b)&& [[\pa_i,\Delta_j],P_{ij}]=0\quad\Leftrightarrow\quad
[\pa_i,\Delta_j]=[\pa_j,\Delta_i]\,,\quad i\neq j\,.\label{2.8}\eea
\end{prop}
\vskip 1mm

\noindent
The final operators that will be introduced in this Section are the
Dunkl's type operators
\beq
D_i=\pa_i+g\Delta_i\,,\qquad g\in\RR\,,
\label{2.9}\eeq
which are the differential-permutation operators acting on the
function space $f(x_1,x_2,\ldots,x_N)$ $\in C^\infty(\RR^N)$.
Let us now prove the following Theorem about the algebraic
relations for the operators $P_{ij}$, $D_k$, and $x_l$.
\begin{theorem}\label{t2.3}
\bea
(a)&& P_{ij}D_k=D_kP_{ij}\qquad\mbox{if}\qquad\{k\}\cap\{i,j\}
=\emptyset\,,\nonumber\\
&&\, P_{ij}D_j=D_iP_{ij}\,,\label{2.10}\\
(b)&& [D_i,D_j]=0\,,\label{2.11}\\
(c)&& [D_i,x_j]=\delta_{ij}(\id+g\sum_{k\neq i}P_{ik})-(1-\delta_{ij})gP_{ij}
\,.\label{2.12}\eea\end{theorem}
\Proof The statement $(a)$ follows from (\ref{2.9}), (\ref{2.7}), (\ref{2.4}).
The commutativity of the operators $D_i$ results from
(\ref{2.9}), (\ref{2.8}), (\ref{2.7}), (\ref{2.5}). The last commutator
$(c)$ is easy to derive from the relation (\ref{2.6}).
\vskip 1mm

The commutative operators $D_i$ were first introduced and studied in
\cite{D}.
%
%
\section{Quantum Calogero-Moser system}
\setcounter{equation}{0}
\noindent
Let us define a quantum integrable system by fixing the complete set
of the symmetric polynomials $I_k$, $k=1,\ldots,N$, on $N$ Dunkl's operators
$D_i$ as corresponding mutually commuting integrals of motion
\beq
I_k=\sum_i D_i^k,\qquad k=1,\ldots,N\,.
\label{3.1}\eeq
The first two integrals have the form
\bea
I_1&=&\sum_k\pa_k\,,\nonumber\\
I_2&=&\sum_k\pa_k^2+2g\sum_{i<j}\frac{1}{x_{ij}}\left[\pa_i-\pa_j-
\frac{1}{x_{ij}}(\id-P_{ij})\right]\,.
\nonumber\eea
Introduce now the operation $Res$ which acts on operators sending
symmetric functions to symmetric ones (i.e. operators leaving invariant
the sub-space of symmetric functions) and means the restriction of these
operators on the sub-space of symmetric functions. Then, for instance,
\beq
H_2\equiv Res(I_2)=\sum_k\pa_k^2+2g\sum_{i<j}\frac{1}{x_{ij}}(\pa_i-\pa_j)\,,
\label{3.2}\eeq
since the operator $\id-P_{ij}$ vanishes on symmetric functions.
Define a set of the PDOs $H_k$ of orders from $1$ to $N$ by the rule:
\beq
H_k=Res(I_k)\,,\qquad k=1,\ldots,N\,.
\label{3.3}\eeq
Notice that all these operators (after applying the operation $Res$)
become purely differential operators, i.e. do not contain any $P_{ij}$'s
parts.
\begin{prop}
The operator $H_k$ has the order $k$ and all of them are mutually
commuting
\be [H_i,H_j]=0\,.\ee
\end{prop}
\Proof follows from the relations (\ref{3.3}),(\ref{3.1}),(\ref{2.11}),%
(\ref{2.10}).
\vskip 1mm

\noindent
The set of operators
\be
\widetilde{H_i}=w\circ H_i\circ w^{-1}\,,\qquad w=\prod_{i<j}(x_i-x_j)^g
\ee
gives the integrals of motion of the quantum Calogero-Moser system with
the second integral (Hamiltonian) having the form
\be
\widetilde{H_2}=\sum_{k=1}^N\frac{\pa^2}{\pa x_k^2}-2g(g-1)\sum_{i<j}
\frac{1}{(x_i-x_j)^2}\,.
\ee

We refer here to the lectures \cite{H1,H2} where the Dunkl's type operators
were used for proving the quantum complete integrability of the
(trigonometric) Calogero-Sutherland model and of its generalisation
to the other classical root systems.
%
%
\section{Superstructure}
\setcounter{equation}{0}
\noindent
In this Section we introduce the superintegrability structure for the
quantum Calogero-Moser system which explains the degeneration of this model.
The crucial role in the whole construction is played by the operators
$S_i$, $i=0,1,2,\ldots,$ introduced below.
\begin{prop}\label{t4.1} 
Let $n=1,2,3,\ldots\,.$ Then
\bea
[D_i^n,x_j]&=&\delta_{ij}\left(nD_i^{n-1}+\sum_{k\neq i}R_{n-1}(D_i,D_k)\;
gP_{ik}\right)
\nonumber\\
&&-(1-\delta_{ij})R_{n-1}(D_i,D_j)\; gP_{ij}\,,
\label{4.1}\eea
where $R_{n-1}(x,y)$ is the symmetric polynomial of the order $n-1$
of the form
\be
R_{n-1}(x,y)=\frac{x^n-y^n}{x-y}\,.
\ee
\end{prop}
\Proof is done by induction on $n$ with help of the statement $(c)$
from Theorem \ref{t2.3}.
\vskip 1mm

\noindent
As a corollary of this Proposition we have the following
\begin{prop}\label{t4.2}  
Let $n=1,2,3,\ldots\,,\;\, l=0,1,2,\ldots\,.$ Then
\bea
[x_iD_i^n,x_jD_j^l]&=&\delta_{ij}
\left(nx_iD_i^{l+n-1}+\sum_{k\neq i}x_iD_k^lR_{n-1}(D_i,D_k)\;
gP_{ik}\right)
\nonumber\\
&&-(1-\delta_{ij})x_iD_i^lR_{n-1}(D_i,D_j)\; gP_{ij}\,.
\label{4.2}\eea
\end{prop}
\vskip 1mm

\noindent
Introduce now the additional operators $S_i$ which, together with
the integrals $I_k$, constitute the `superstructure' of the quantum
Calogero-Moser system
\beq
S_k=\sum_{i=1}^N x_i D_i^k\,,\qquad k=0,1,2,\ldots\,.
\label{4.3}\eeq
It is quite straightforward calculation to derive the following
algebra for the operators $S_l$, $I_n$ on the basis of the Propositions
\ref{t4.1} and \ref{t4.2}.
\begin{theorem} \label{t4.3} 
The operators $S_l=\sum_i x_iD_i^l$, $I_n=\sum_i D_i^n$
are closed to the algebra
\bea
(i)&&   [I_l,I_n]=0\,,\label{4.4}\\
(ii)&&  [S_l,I_n]=-nI_{l+n-1}\,,\label{4.5}\\
(iii)&& [S_l,S_n]=(l-n)S_{l+n-1}\,.\label{4.6}\eea
\end{theorem}
\vskip 1mm

\noindent
\begin{defi}
Consider $k$ operators $X_1,\ldots,X_k$. Fix an ordering
$X_1\prec X_2\prec\ldots\prec X_k$. The operators $X_1,\ldots,X_k$
are {\bf algebraically independent} if $p\equiv 0$ is the only polynomial
such that
\be
p(X_1,X_2,\ldots,X_k)=0\,.
\ee\end{defi}
\vskip 1mm

\noindent
For fixed $N$ there are only $N$ additional, algebraically
independent operators $S_i$ because of the easily verified identity
\beq
S_{N+k}=\sum_{i=1}^N (-1)^{i+1} S_{N+k-i}\;M_i(D)\,,\qquad
k=0,1,2,\ldots\,,
\label{4.7}\eeq
where $M_i(D)$ are elementary monomial symmetric functions
\beq
M_i(D)=\sum_{k_1<\ldots<k_i}D_{k_1}\cdot\ldots\cdot D_{k_i}\,.
\label{4.8}\eeq
Let us think always of the first $N$ lower order operators
\be
S_0\,,\quad S_1\,,\quad \ldots\,, \quad S_{N-1}
\ee
as $N$ chosen algebraically independent
operators additional to the $I_k$, $k=1,\ldots,N$.

It is easy to observe that the new operators $S_i$ commute with all
the permutations
\be
[S_i,P_{jk}]=0\,,
\ee
(as well as the $I_n$ do!). Hence, they send symmetric functions
to symmetric ones and one can restrict the relations (\ref{4.4})--%
(\ref{4.6}) on the sub-space of the symmetric functions. So, we get
the following algebra ($Res(I_n)=H_n$):
\bea
           {[H_l,H_n]}&=&0\,,\label{4.9}\\
      {[Res(S_l),H_n]}&=&-nH_{l+n-1}\,,\label{4.10}\\
 {[Res(S_l),Res(S_n)]}&=&(l-n)Res(S_{l+n-1})\,.\label{4.11}\eea
The operators $Res(S_l)$ and $H_n$ are now purely differential operators
of orders $l$ and $n$, respectively. Let us construct the action of the
operators $Res(S_l)$ on the common eigenfunction $\Psi_{\vec m}(\vec x)$
of the quantum integrals of motion $H_k$
\beq
H_k\;\Psi_{\vec m}(\vec x)=\sum_{i=1}^N m_i^k \;\Psi_{\vec m}(\vec x)\,,
\qquad m_i\in \RR\,.
\label{4.12}\eeq
The eigenfunction $\Psi_{\vec m}(\vec x)\in C^\infty(\RR^{2N})$ depends
on $N$ variables $\vec{x}=(x_1,\ldots,x_N)$
and on $N$ real spectral parameters $\vec{m}=(m_1,\ldots,m_N)$.
\begin{prop}\label{t4.5}
The multiplication operators
\beq
{\cal H}_n \;\Psi_{\vec{m}} = \sum_{i=1}^N m_i^n \;\Psi_{\vec m}
\label{4.13}\eeq
and the following first-order differential operators (in $m_i$'s!)
\beq
{\cal S}_l \;\Psi_{\vec m} = - \sum_{i=1}^N m_i^l \frac{\pa}{\pa m_i}\;
\Psi_{\vec m}
\label{4.14}\eeq
give a representation of the algebra (\ref{4.9})--(\ref{4.11}), i.e.
\bea
      {[{\cal H}_l,{\cal H}_n]}&=&0\,,\label{4.15}\\
      {[{\cal S}_l,{\cal H}_n]}&=&-n{\cal H}_{l+n-1}\,,\label{4.16}\\
      {[{\cal S}_l,{\cal S}_n]}&=&(l-n){\cal S}_{l+n-1}\,.\label{4.17}\eea
\end{prop}
\Proof can be done by straightforward computation.
\vskip 1mm

\noindent
It is well-known \cite{O,J} that the so-called multi-variable Bessel
function $J^{(g)}_{\vec m}(\vec x)$ solves the spectral problem (\ref{4.12})
\beq
H_k\;J^{(g)}_{\vec m}(\vec x)=\sum_{i=1}^N m_i^k \;J^{(g)}_{\vec m}(\vec x)\,,
\qquad m_i\in \RR\,.
\label{4.18}\eeq
Hence, as the corollary of the relations (\ref{4.13})--(\ref{4.17})
we have the following identities for such functions:
\beq
Res\left[\sum_{i=1}^Nx_i\left(\frac{\pa}{\pa x_i}
+g\sum_{j\neq i}\frac{1}{x_i-x_j}(\id-P_{ij})\right)^l\,\right]\;
J^{(g)}_{\vec m}(\vec x)=\sum_{i=1}^N m_i^l \frac{\pa}{\pa m_i}\;
J^{(g)}_{\vec m}(\vec x)\,.
\label{4.19}\eeq
Here in the l.h.s. we have the order $l$ differential operator in $x_i$'s
while in the r.h.s. we have the first-order differential operator in
the spectral parameters $m_i$'s. In the limit $g=0$ the multi-variable
Bessel function $J^{(g)}_{\vec m}(\vec x)$ turns into the
symmetrised exponential $\sum_{\sigma_x\in\SS_N}\sigma_x(
\exp(\vec{m}\vec{x}))$ and the relations (\ref{4.19}) look like
($l=0,1,2,\ldots$)
\beq
 \sum_{i=1}^Nx_i\frac{\pa^l}{\pa x_i^l}\;
 \sum_{\sigma_x\in\SS_N}\sigma_x(\exp(\vec{m}\vec{x}))
=\sum_{i=1}^Nm_i^l\frac{\pa}{\pa m_i}\;
\sum_{\sigma_x\in\SS_N}\sigma_x(\exp(\vec{m}\vec{x}))\,.
\label{4.20}\eeq
%
%
\section{Quadratic algebra}
\setcounter{equation}{0}
\noindent
Introduce the additional operators
\beq
A^{(k)}_{ij}=S_iI_{j+k-1}-S_jI_{i+k-1}\,,
\label{5.1}\eeq
where $k=1,2,3,\ldots\,,\;i,j=0,1,2,\ldots\,.$ By definition we have the
following properties:
\be
A^{(k)}_{ii}=0\,,\qquad A^{(k)}_{ij}=-A^{(k)}_{ji}\,,
\ee
so that we assume that $i<j$. The order of the PDO which is the restriction
of the operator $A^{(k)}_{ij}$ is
\beq
\mbox{order}[Res(A^{(k)}_{ij})]=i+j+k-1\,.
\label{5.2}\eeq
\begin{prop}
\bea
(a)&& [A^{(k)}_{ij},S_n]=iA^{(k)}_{i+n-1,j}+jA^{(k)}_{i,j+n-1}\nonumber\\
 &&\qquad\qquad +(k-1)A^{(k+n-1)}_{ij}-nA^{(k+1-n)}_{i+n-1,j+n-1}
\,,\label{5.3}\\
(b)&& [A^{(k)}_{ij},I_k]=0\,,\label{5.4}\\
(c)&& [Res(A^{(k)}_{ij}),H_k]=0\,.
\label{5.5}\eea
\end{prop}
\Proof follows from the definition (\ref{5.1}) and from the Theorem
\ref{t4.3}.
\vskip 1mm

\noindent
Notice that from the statement $(c)$ it follows that we have constructed
a big supply of the PDOs $Res(A^{(k)}_{ij})$, $i,j=0,1,2,\ldots,$ all commuting
to the integral $H_k$. These additional operators of the order $i+j+k-1$
describe the hidden symmetry (or the super-integrability)
of the quantum integrals of motion for the Calogero-Moser system.

We are now in a position to derive the global quadratic algebra
for the integrals $I_n$ and the `hidden symmetry generators' $A^{(k)}_{ij}$.
\begin{theorem}\label{t5.2} 
The operators $I_n$ and $A^{(k)}_{ij}$ are closed to the
quadratic algebra
\bea
(a)&& [I_i,I_j]=0\,,\label{5.6}\\
(b)&& [A^{(k)}_{ij},I_n]=-n(I_{i+n-1}I_{j+k-1}-I_{j+n-1}I_{i+k-1})
\,,\label{5.7}\\
(c)&& [A^{(k)}_{ij},A^{(k')}_{i'j'}]=
    i(A^{(k)}_{i+i'-1,j}I_{j'+k'-1}-A^{(k)}_{i+j'-1,j}I_{i'+k'-1})\nonumber\\
&&\qquad\qquad\quad
+i'(A^{(k')}_{i'+j-1,j'}I_{i+k-1}-A^{(k')}_{i'+i-1,j'}I_{j+k-1})\nonumber\\
&&\qquad\qquad\quad
+ j(A^{(k)}_{j+j'-1,i}I_{i'+k'-1}-A^{(k)}_{j+i'-1,i}I_{j'+k'-1})\nonumber\\
&&\qquad\qquad\quad
+j'(A^{(k')}_{j'+i-1,i'}I_{j+k-1}-A^{(k')}_{j'+j-1,i'}I_{i+k-1})\nonumber\\
&&\qquad\qquad\quad
+(k-1)(A^{(k-1+i')}_{ij}I_{k'-1+j'}-A^{(k-1+j')}_{ij}I_{k'-1+i'})\nonumber\\
&&\qquad\qquad\quad
+(k'-1)(A^{(k'-1+j)}_{i'j'}I_{k-1+i}-A^{(k'-1+i)}_{i'j'}I_{k-1+j})\,.
\label{5.8}\eea
\end{theorem}
\vskip 1mm

\noindent
Not all of the operators $A^{(k)}_{ij}$ are algebraically independent.
Actually, for each given $k$ (and a fixed $N$) we can supply only $N-1$
operators $A^{(k)}_{ij}$ which all commute to the $I_k$ and all of them
plus the integrals of motion are algebraically independent.
In this way we get a $N\times N$ table putting, for instance,
$i=0$ to diminish the order of the generators:
$$
\left[\matrix{I_1    & A^{(1)}_{01} &  \cdots & A^{(1)}_{0,N-1} \cr
              I_2    & A^{(2)}_{01} &  \cdots & A^{(2)}_{0,N-1} \cr
              \vdots & \vdots    &  \ddots & \vdots         \cr
              I_N    & A^{(N)}_{01} &  \cdots & A^{(N)}_{0,N-1}  }\right].
$$
If we take the restriction of all operators in the table to the
invariant sub-space of the symmetric functions and denote
$B^{(k)}_j=Res(A^{(k)}_{0j})$ then we have the same table that was
announced in the Introduction Section:
\beq
\left[\matrix{H_1    & B^{(1)}_1 &  \cdots & B^{(1)}_{N-1} \cr
              H_2    & B^{(2)}_1 &  \cdots & B^{(2)}_{N-1} \cr
              \vdots & \vdots    &  \ddots & \vdots         \cr
              H_N    & B^{(N)}_1 &  \cdots & B^{(N)}_{N-1}  }\right].
\label{table}\eeq
It is easy now to derive the quadratic algebra between the entries of this
table.
\begin{theorem}\label{t5.3} 
The operators $H_n$ and $B^{(k)}_{j}=Res(A^{(k)}_{0j})$ are closed to the
quadratic algebra
\bea
(a)&& [H_i,H_j]=0\,,\label{5.9}\\
(b)&& [B^{(k)}_{j},H_n]=-n(H_{n-1}H_{j+k-1}-H_{k-1}H_{j+n-1})
\,,\label{5.10}\\
(c)&& [B^{(k)}_{j},B^{(k')}_{j'}]=
j(-B^{(k)}_{j+j'-1}H_{k'-1}+B^{(k)}_{j-1}H_{j'+k'-1})\nonumber\\
&&\qquad\qquad\quad
+j'(-B^{(k')}_{j'-1}H_{j+k-1}+B^{(k')}_{j'+j-1}H_{k-1})\nonumber\\
&&\qquad\qquad\quad
+(k-1)(B^{(k-1)}_{j}H_{k'-1+j'}-B^{(k-1+j')}_{j}H_{k'-1})\nonumber\\
&&\qquad\qquad\quad
+(k'-1)(B^{(k'-1+j)}_{j'}H_{k-1}-B^{(k'-1)}_{j'}H_{k-1+j})\,.
\label{5.11}\eea
\end{theorem}
\vskip 1mm

\noindent
Recall \cite{W} that the superintegrability of the classical Calogero-Moser
system means that each $H_n$ of the integrals of motion is in involution with
another $2N-2$ quantities. In the quantum case we have shown (cf. the
statements $(a)$ and $(b)$ in the Theorem \ref{t5.3}) that the $H_n$
commutes with $2N-2$
operators: $H_1,\ldots,H_{n-1},H_{n+1},\ldots,H_{N},\;B^{(n)}_1,\ldots,
B^{(n)}_{N-1}$. This is the fact of the
quantum superintegrability or degeneration
of the quantum integrable system.
\vskip 1mm

\noindent
{\bf Remark} For the table (\ref{table}) one gets in the r.h.s.'s
of the commutators (\ref{5.10})--(\ref{5.11}) some $H_k$'s and $B^{(k)}_j$'s
which are out of the table. The simple analysis shows that all these operators
are not algebraically independent from those in the table, i.e. they can be
expressed as some polynomials (linear in $B^{(k)}_j$'s)
on the operators inside the table.
\vskip 1mm

\noindent
Let us consider in details two first examples: $N=2$ and $N=3$.
%
%
\section{$N=2$ case}
\setcounter{equation}{0}
\noindent
We have the following $2\times 2$ table in the case of $N=2$:
\be
\left[\matrix{H_1    & B^{(1)}_1 \cr
              H_2    & B^{(2)}_1 }\right],
\ee
and the quadratic algebra of the form
\bea
[H_1,H_2]&=&[H_1,B^{(1)}_1]=[H_2,B^{(2)}_1]=0\,,\label{6.1}\\
{[B^{(1)}_1,H_2]}&=&-2(H_1^2-2H_2)\,,\label{6.2}\\
{[B^{(2)}_1,H_1]}&=&H_1^2-2H_2\,,\label{6.3}\\
{[B^{(1)}_1,B^{(2)}_1]}&=&4B^{(2)}_1-2B^{(1)}_1H_1\,.\label{6.4}
\eea
Let us denote the restrictions of the operators $S_i$ by the small
letters:
\beq s_i=Res(S_i)\,.\label{6.5}\eeq
Then the additional operators $s_0,s_1,s_2$ satisfy the following relations:
\bea
{[s_0,H_1]}&=&-2\cdot\id\,,
\qquad [s_0,B^{(1)}_1]=0\,,\label{6.6}\\
{[s_0,H_2]}&=&-2H_1\,,\qquad [s_0,B^{(2)}_1]=-B^{(1)}_1\,,\label{6.7}\\
{[s_1,H_1]}&=&-H_1\,,\qquad \;\, [s_1,B^{(1)}_1]=0\,,\label{6.8}\\
{[s_1,H_2]}&=&-2H_2\,,\qquad [s_1,B^{(2)}_1]=-B^{(2)}_1\,,\qquad
[s_0,s_1]=-s_0\,,\label{6.9}\\
{s_2}&=&s_1H_1+\frac12s_0(H_2-H_1^2)\,,\label{6.10}\\
{[s_2,H_1]}&=&-H_2\,,\qquad \qquad\;\;[s_2,B^{(1)}_1]=B^{(1)}_1H_1-2B^{(2)}_1
\,,\nonumber\\
{[s_2,H_2]}&=&H_1^3-3H_1H_2\,,\quad  [s_2,B^{(2)}_1]=
\frac32 B^{(1)}_1(H_1^2-H_2)-2B^{(2)}_1H_1\,.\nonumber
\eea
The explicit form of all these operators is as follows:
\bea
H_1&=&\pa_1+\pa_2\,,\label{6.11}\\
H_2&=&\pa_1^2+\pa_2^2+\frac{2g}{x_1-x_2}(\pa_1-\pa_2)\,,\label{6.12}\\
{B^{(1)}_1}&=&(x_1-x_2)(\pa_2-\pa_1)\,,\label{6.13}\\
{B^{(2)}_1}&=&x_2\pa_1^2+x_1\pa_2^2-(x_1+x_2)\pa_1\pa_2+
2g\frac{x_1+x_2}{x_1-x_2}(\pa_1-\pa_2)\,,\label{6.14}\\
s_0&=&x_1+x_2\,,\label{6.15}\\
s_1&=&x_1\pa_1+x_2\pa_2\,,\label{6.16}\\
s_2&=&x_1\pa_1^2+x_2\pa_2^2+g\frac{x_1+x_2}{x_1-x_2}(\pa_1-\pa_2)\,.
\label{6.17}\eea
\begin{theorem}
The spectral problem
\bea
{H_1\;J^{(g)}_{m_1m_2}(x_1,x_2)}&=&(m_1+m_2)\;J^{(g)}_{m_1m_2}(x_1,x_2)\,,
\qquad m_1,m_2\in\RR\,,\label{6.18}\\
{H_2\;J^{(g)}_{m_1m_2}(x_1,x_2)}&=&(m_1^2+m_2^2)\;J^{(g)}_{m_1m_2}(x_1,x_2)
\label{6.19}\eea
has the following symmetric function as a solution
\bea
&&{J^{(g)}_{m_1m_2}(x_1,x_2)}=\e^{\frac{m_+}{2}x_+}\,(m_-x_-)^{
\frac12-g}\;I_{g-\frac12}\left(\frac{m_-}{2}x_-\right)\,,
\label{6.20}\\
&&x_\pm=x_1\pm x_2\,,\qquad m_\pm=m_1\pm m_2\,,\nonumber
\eea
where
\beq
I_\nu(z)=\sum_{k=0}^\infty \frac{(\frac{z}{2})^{2k+\nu}}{k!\Gamma(k+\nu+1)}
\label{6.21}\eeq
is the modified Bessel function of the first kind \cite[7.2(12)]{E}.
Moreover
\beq
J^{(0)}_{m_1m_2}(x_1,x_2)=\frac{1}{\sqrt{\pi}}
(\e^{m_1x_1+m_2x_2}+\e^{m_1x_2+m_2x_1})\,.
\label{6.22}\eeq
\end{theorem}
Notice that if $\nu\in\RR$ and $z>0$ then $I_\nu(z)\in\RR$.
\vskip 1mm

\noindent
\Proof can be done through the separation of variables $x_+$ and
$x_-$ in the spectral problem (\ref{6.18})--(\ref{6.19}).
\vskip 1mm

\noindent
\begin{theorem}
The action of the operators (\ref{6.11})--(\ref{6.17}) on the
eigenfunction (\ref{6.20}) is as follows:
\bea
{H_1\;J^{(g)}_{m_1m_2}(x_1,x_2)}&=&
(m_1+m_2)\;J^{(g)}_{m_1m_2}(x_1,x_2)\,,\label{6.23}\\
{H_2\;J^{(g)}_{m_1m_2}(x_1,x_2)}&=&
(m_1^2+m_2^2)\;J^{(g)}_{m_1m_2}(x_1,x_2)\,,\label{6.24}\\
{B^{(1)}_1\;J^{(g)}_{m_1m_2}(x_1,x_2)}&=&
(m_2-m_1)(\pa_{m_1}-\pa_{m_2})\;J^{(g)}_{m_1m_2}(x_1,x_2)\,,\label{6.25}\\
{B^{(2)}_1\;J^{(g)}_{m_1m_2}(x_1,x_2)}&=&
(m_2-m_1)(m_2\pa_{m_1}-m_1\pa_{m_2})\;J^{(g)}_{m_1m_2}(x_1,x_2)
\,,\label{6.26}\\
s_0\;J^{(g)}_{m_1m_2}(x_1,x_2)&=&
(\pa_{m_1}+\pa_{m_2})\;J^{(g)}_{m_1m_2}(x_1,x_2)\,,\label{6.27}\\
s_1\;J^{(g)}_{m_1m_2}(x_1,x_2)&=&
(m_1\pa_{m_1}+m_2\pa_{m_2})\;J^{(g)}_{m_1m_2}(x_1,x_2) \,,\label{6.28}\\
s_2\;J^{(g)}_{m_1m_2}(x_1,x_2)&=&
(m_1^2\pa_{m_1}+m_2^2\pa_{m_2})\;J^{(g)}_{m_1m_2}(x_1,x_2) \,.
\label{6.29}\eea
\end{theorem}
\Proof follows from the Proposition \ref{t4.5} (cf. formula (\ref{4.19}))
and the definition
of $A^{(k)}_{ij}$ (\ref{5.1}) (recall also that $B^{(k)}_{j}=$
$Res(A^{(k)}_{0j})$).
%
%
\section{$N=3$ case}
\setcounter{equation}{0}
\noindent
We have the following $3\times 3$ table in the case of $N=3$:
\be
\left[\matrix{H_1    & B^{(1)}_1& B^{(1)}_2 \cr
              H_2    & B^{(2)}_1& B^{(2)}_2 \cr
              H_3    & B^{(3)}_1& B^{(3)}_2 }\right],
\ee
and the non-linear algebra of the form
\beq [H_i,H_j]=[H_i,B^{(i)}_1]=[H_i,B^{(i)}_2]=0\,,\label{77.1}\eeq
\bea
{[H_1,B^{(2)}_1]}&=&3H_2-H_1^2\,,\label{7.1}\\
{[H_1,B^{(2)}_2]}&=&3H_3-H_1H_2\,,\label{7.2}\\
{[H_1,B^{(3)}_1]}&=&3H_3-H_1H_2\,,\label{7.3}\\
{[H_1,B^{(3)}_2]}&=&4H_3H_1+\frac12 H_2^2-3H_2H_1^2+\frac12 H_1^4\,,
\label{7.4}\eea
\bea
{[H_2,B^{(1)}_1]}&=&-2(3H_2-H_1^2)\,,\label{7.5}\\
{[H_2,B^{(1)}_2]}&=&-2(3H_3-H_1H_2)\,,\label{7.6}\\
{[H_2,B^{(3)}_1]}&=&-2(H_2^2-H_1H_3)\,,\label{7.7}\\
{[H_2,B^{(3)}_2]}&=&
-2H_3H_2+\frac83 H_3H_1^2+H_2^2H_1-2H_2H_1^3+\frac13 H_1^5\,,\label{7.8}
\eea
\bea
{[H_3,B^{(1)}_1]}&=&-3(3H_3-H_1H_2)\,,\label{7.9}\\
{[H_3,B^{(1)}_2]}&=&-12H_3H_1-\frac32 H_2^2+9H_2H_1^2-\frac32 H_1^4\,,
\label{7.10}\\
{[H_3,B^{(2)}_1]}&=&3(H_2^2-H_1H_3)\,,\label{7.11}\\
{[H_3,B^{(2)}_2]}&=&
3H_3H_2-4H_3H_1^2-\frac32 H_2^2H_1+3H_2H_1^3-\frac12 H_1^5\,,\label{7.12}
\eea
\bea
{[B^{(1)}_1,B^{(1)}_2]}&=&3B^{(1)}_2-2B^{(1)}_1H_1\,,\label{7.13}\\
{[B^{(1)}_1,B^{(2)}_1]}&=&6B^{(2)}_1-2B^{(1)}_1H_1\,,\label{7.14}\\
{[B^{(1)}_1,B^{(2)}_2]}&=&9B^{(2)}_2-2(B^{(1)}_2+B^{(2)}_1)H_1\,,\label{7.15}\\
{[B^{(1)}_1,B^{(3)}_1]}&=&9B^{(3)}_1-2B^{(2)}_1H_1-B^{(1)}_1H_2\,,\label{7.16}\\
{[B^{(1)}_1,B^{(3)}_2]}&=&
12B^{(3)}_2-2(B^{(3)}_1+B^{(2)}_2)H_1-B^{(1)}_2H_2\,,\label{7.17}\\
{[B^{(1)}_2,B^{(2)}_1]}&=&
3(B^{(3)}_1+B^{(2)}_2)-2B^{(1)}_2H_1+B^{(1)}_1H_2\,,\label{7.18}\\
{[B^{(1)}_2,B^{(2)}_2]}&=&
3B^{(3)}_2+6B^{(2)}_2H_1+B^{(2)}_1(H_2-3H_1^2)\nonumber\\
&&-B^{(1)}_2(H_2+2H_1^2)+B^{(1)}_1(H_1^3-H_1H_2+2H_3)\,,\label{7.19}\\
{[B^{(1)}_2,B^{(3)}_1]}&=&
3B^{(3)}_2+6B^{(3)}_1H_1+B^{(2)}_1(H_2-3H_1^2)\nonumber\\
&&-2B^{(1)}_2H_2+B^{(1)}_1(H_1^3-3H_1H_2+4H_3)\,,\label{7.20}\\
{[B^{(1)}_2,B^{(3)}_2]}&=&
12B^{(3)}_2H_1+(B^{(3)}_1+B^{(2)}_2)(H_2-3H_1^2)\nonumber\\
&&+B^{(1)}_2(H_1^3-5H_1H_2+2H_3)+
B^{(1)}_1(\frac13 H_1^3-H_1H_2+\frac83 H_3)H_1\,,\label{7.21}\\
{[B^{(2)}_1,B^{(2)}_2]}&=&
(2B^{(2)}_2-B^{(3)}_1)H_1-(2B^{(2)}_1+B^{(1)}_2)H_2+B^{(1)}_1H_3
\,,\label{7.22}\\
{[B^{(2)}_1,B^{(3)}_1]}&=&
3B^{(3)}_1H_1-4B^{(2)}_1H_2+B^{(1)}_1H_3\,,\label{7.23}\\
{[B^{(2)}_1,B^{(3)}_2]}&=&
4B^{(3)}_2H_1-3(B^{(3)}_1+B^{(2)}_2)H_2+
B^{(1)}_1(\frac16 H_1^4-H_2H_1^2+\frac12 H_2^2+\frac43 H_3H_1)\,,\label{7.24}\\
{[B^{(2)}_2,B^{(3)}_1]}&=&
B^{(3)}_2H_1+2B^{(3)}_1H_1^2-3B^{(2)}_2H_2+B^{(2)}_1(H_2-H_1^2)H_1\nonumber\\
&&+B^{(1)}_2H_3+\frac13 B^{(1)}_1(2H_3-3H_1H_2+H_1^3)H_1\,,\label{7.25}\\
{[B^{(2)}_2,B^{(3)}_2]}&=&
B^{(3)}_2(4H_1^2-H_2)+B^{(3)}_1(H_1H_2-H_1^3-2H_3)
-B^{(2)}_2(H_1H_2+H_1^3+2H_3)\label{7.26}\\
&&+B^{(2)}_1(\frac83 H_3-H_1H_2+\frac13 H_1^3)H_1
+B^{(1)}_2(2H_3H_1-2H_1^2H_2+\frac12 H_1^4+\frac12 H_2^2)\,,\nonumber\\
{[B^{(3)}_1,B^{(3)}_2]}&=&
3B^{(3)}_2H_2-2B^{(3)}_1(H_3+H_1H_2)-2B^{(2)}_2H_3\nonumber\\
&&+B^{(2)}_1(\frac13 H_1^3-H_1H_2+\frac83 H_3)H_1-\frac13
B^{(1)}_1(H_1^3-3H_1H_2+2H_3)H_2\,.\label{7.27}
\eea
The additional operators $s_0,s_1,s_2$ satisfy the following relations:
\bea
{[s_0,H_1]}&=&-3\cdot\id\,,\qquad [s_1,H_1]=-H_1\,,\label{7.28}\\
{[s_0,H_2]}&=&-2H_1\,,\qquad [s_1,H_2]=-2H_2\,,\label{7.29}\\
{[s_0,H_3]}&=&-3H_2\,,\qquad [s_1,H_3]=-3H_3\,,\label{7.30}\eea
\bea
{[s_2,H_1]}&=&-H_2\,,\label{7.31}\\
{[s_2,H_2]}&=&-2H_3\,,\label{7.32}\\
{[s_2,H_3]}&=&-\frac12 H_1^4+3H_2H_1^2-\frac32 H_2^2-4H_3H_1\,,\label{7.33}\eea
\beq
[s_0,s_1]=-s_0\,,\qquad [s_0,s_2]=-2s_1\,,\qquad [s_1,s_2]=-s_2\,,
\label{7.34}\eeq
\bea
{[s_0,B^{(1)}_1]}&=&0\,,\qquad\qquad\;\;[s_0,B^{(2)}_2]=-2B^{(2)}_1-B^{(1)}_2
\,,\label{7.35}\\
{[s_0,B^{(1)}_2]}&=&-2B^{(1)}_1\,,\qquad [s_0,B^{(3)}_1]=-2B^{(2)}_1
\,,\label{7.36}\\
{[s_0,B^{(2)}_1]}&=&-B^{(1)}_1\,,\;\;
\qquad [s_0,B^{(3)}_2]=-2B^{(3)}_1-2B^{(2)}_2
\,,\label{7.37}\eea
\bea
{[s_1,B^{(1)}_1]}&=&0\,,\qquad\qquad\;\;[s_1,B^{(2)}_2]=-2B^{(2)}_2
\,,\label{7.38}\\
{[s_1,B^{(1)}_2]}&=&-B^{(1)}_2\,,\qquad\;\;[s_1,B^{(3)}_1]=-2B^{(3)}_1
\,,\label{7.39}\\
{[s_1,B^{(2)}_1]}&=&-B^{(2)}_1\,,\;\;
\qquad [s_1,B^{(3)}_2]=-3B^{(3)}_2
\,,\label{7.40}\eea
\beq
[s_2,B^{(1)}_1]=B^{(1)}_2-2B^{(2)}_1\,,\qquad
[s_2,B^{(1)}_2]=-2B^{(3)}_1\,,\qquad
[s_2,B^{(2)}_1]=B^{(2)}_2-3B^{(3)}_1\,,\label{7.41}\eeq
\bea
{[s_2,B^{(2)}_2]}&=&-B^{(3)}_2-2B^{(3)}_1H_1-B^{(2)}_1(H_2-H_1^2)
 -\frac13 B^{(1)}_1(2H_3-3H_1H_2+H_1^3) \,,\label{7.42}\\
{[s_2,B^{(3)}_1]}&=&B^{(3)}_2-4B^{(3)}_1H_1-2B^{(2)}_1(H_2-H_1^2)
-\frac23 B^{(1)}_1(2H_3-3H_1H_2+H_1^3) \,,\label{7.43}\\
{[s_2,B^{(3)}_2]}&=&-2B^{(3)}_2H_1+B^{(3)}_1(H_2-\frac53 H_1^2)
-B^{(2)}_2(H_2-H_1^2) +B^{(2)}_1(-\frac83 H_3+\frac23 H_1H_2+\frac23 H_1^3)
\nonumber\\
&&-\frac13 B^{(1)}_2(2H_3-3H_1H_2+H_1^3)
-\frac13 B^{(1)}_1(H_1^2-H_2)(H_1^2-2H_2)\,.\label{7.44}
\eea
The explicit form of all these operators is as follows:
\bea
H_1&=&\pa_1+\pa_2+\pa_3\,,\label{7.45}\\
H_2&=&\pa_1^2+\pa_2^2+\pa_3^2+\frac{2g}{x_{12}}\pa_{12}
+\frac{2g}{x_{13}}\pa_{13}+\frac{2g}{x_{23}}\pa_{23}
\,,\label{7.46}\\
H_3&=&\pa_1^3+\pa_2^3+\pa_3^3
+3g\left(\frac{1}{x_{12}}+\frac{1}{x_{13}}\right)\pa_1^2
+3g\left(\frac{1}{x_{21}}+\frac{1}{x_{23}}\right)\pa_2^2
+3g\left(\frac{1}{x_{31}}+\frac{1}{x_{32}}\right)\pa_3^2\nonumber\\
&&+\frac{6g^2}{x_{12}x_{13}}\pa_1+\frac{6g^2}{x_{21}x_{23}}\pa_2
+\frac{6g^2}{x_{31}x_{32}}\pa_3
\,,\label{7.47}\\
{B^{(1)}_1}&=&(x_{21}+x_{31})\pa_1+(x_{12}+x_{32})\pa_2+(x_{13}+x_{23})\pa_3
\,,\label{7.48}\\
{B^{(2)}_1}&=&(x_2+x_3)(\pa_1^2-\pa_2\pa_3)+(x_1+x_3)(\pa_2^2-\pa_1\pa_3)+
(x_1+x_2)(\pa_3^2-\pa_1\pa_2)\nonumber\\&&
+2g(x_1+x_2+x_3)\left[\left(\frac{1}{x_{12}}+\frac{1}{x_{13}}\right)\pa_1
+\left(\frac{1}{x_{21}}+\frac{1}{x_{23}}\right)\pa_2
+\left(\frac{1}{x_{31}}+\frac{1}{x_{32}}\right)\pa_3\right]
\,,\label{7.49}\\
{B^{(1)}_2}&=&(x_{21}+x_{31})\left(\pa_1^2-\frac{g}{x_{23}}\pa_{23}\right)
+(x_{12}+x_{32})\left(\pa_2^2-\frac{g}{x_{13}}\pa_{13}\right)\nonumber\\&&
+(x_{13}+x_{23})\left(\pa_3^2-\frac{g}{x_{12}}\pa_{12}\right)
\,,\label{7.50}\\
{B^{(2)}_2}&=&s_0H_3-s_2H_1\,,\qquad B^{(3)}_1=s_0H_3-s_1H_2\,,\label{7.51}\\
{B^{(3)}_2}&=&s_0(\frac16 H_1^4-H_2H_1^2+\frac12 H_2^2+\frac43H_1H_3)
-s_2H_2\,,\label{7.52}\eea
where
\bea
s_0&=&x_1+x_2+x_3\,,\label{7.53}\\
s_1&=&x_1\pa_1+x_2\pa_2+x_3\pa_3\,,\label{7.54}\\
s_2&=&x_1\pa_1^2+x_2\pa_2^2+x_3\pa_3^2
+g\frac{x_1+x_2}{x_{12}}\pa_{12}+g\frac{x_1+x_3}{x_{13}}\pa_{13}
+g\frac{x_2+x_3}{x_{23}}\pa_{23}\,,\label{7.55}\\
s_3&=&s_2H_1+\frac12 s_1(H_2-H_1^2)+s_0(\frac13 H_3-\frac12 H_1H_2+
\frac16 H_1^3)\,. \label{7.56}\eea
Here we use the notations $x_{ij}=x_i-x_j$ (the same is for $m_{ij}$ below)
and $\pa_{ij}=\pa_i-\pa_j$.
\begin{theorem}
The action of the operators (\ref{7.45})--(\ref{7.55}) on the common symmetric
eigenfunction $J^{(g)}_{m_1m_2m_3}(x_1,x_2,x_3)$ of the commuting operators
$H_1,H_2,H_3$ is as follows:
\bea
{H_1\;J^{(g)}_{m_1m_2m_3}(x_1,x_2,x_3)}&=&
(m_1+m_2+m_3)\;J^{(g)}_{m_1m_2m_3}(x_1,x_2,x_3)\,,
\qquad m_1,m_2,m_3\in\RR\,,\label{7.57}\\
{H_2\;J^{(g)}_{m_1m_2m_3}(x_1,x_2,x_3)}&=&
(m_1^2+m_2^2+m_3^2)\;J^{(g)}_{m_1m_2m_3}(x_1,x_2,x_3)\,,\label{7.58}\\
{H_3\;J^{(g)}_{m_1m_2m_3}(x_1,x_2,x_3)}&=&
(m_1^3+m_2^3+m_3^3)\;J^{(g)}_{m_1m_2m_3}(x_1,x_2,x_3)\,,\label{7.59}\\
{B^{(1)}_1\;J^{(g)}_{m_1m_2m_3}(x_1,x_2,x_3)}&=&
[(m_{21}+m_{31})\pa_{m_1}+(m_{12}+m_{32})\pa_{m_2}
\nonumber\\&&
+(m_{13}+m_{23})\pa_{m_3}]\;J^{(g)}_{m_1m_2m_3}(x_1,x_2,x_3)\,,\label{7.60}\\
{B^{(1)}_2\;J^{(g)}_{m_1m_2m_3}(x_1,x_2,x_3)}&=&
[(m_{3}^2+m_{2}^2-2m_{1}^2)\pa_{m_1}
+(m_{1}^2+m_{3}^2-2m_{2}^2)\pa_{m_2}\nonumber\\&&
+(m_{1}^2+m_{2}^2-2m_{3}^2)\pa_{m_3}]
\;J^{(g)}_{m_1m_2m_3}(x_1,x_2,x_3)\,,\label{7.61}\\
{B^{(2)}_1\;J^{(g)}_{m_1m_2m_3}(x_1,x_2,x_3)}&=&
 [(m_{1}^2-m_{2}m_{3})(\pa_{m_2}+\pa_{m_3})
+(m_{2}^2-m_{1}m_{3})(\pa_{m_1}+\pa_{m_3})\nonumber\\&&
+(m_{3}^2-m_{1}m_{2})(\pa_{m_1}+\pa_{m_2})]
\;J^{(g)}_{m_1m_2m_3}(x_1,x_2,x_3)\,,\label{7.62}\\
{B^{(2)}_2\;J^{(g)}_{m_1m_2m_3}(x_1,x_2,x_3)}&=&
[(m_{2}^3+m_{3}^3-m_{1}^2(m_2+m_3))\pa_{m_1}
+(m_{1}^3+m_{3}^3-m_{2}^2(m_1+m_3))\pa_{m_2}\nonumber\\&&
+(m_{1}^3+m_{2}^3-m_{3}^2(m_1+m_2))\pa_{m_3}]
\;J^{(g)}_{m_1m_2m_3}(x_1,x_2,x_3)\,,\label{7.63}\\
{B^{(3)}_1\;J^{(g)}_{m_1m_2m_3}(x_1,x_2,x_3)}&=&
 [(m_{2}^3+m_{3}^3-m_{1}(m_2^2+m_3^2))\pa_{m_1}
+(m_{1}^3+m_{3}^3-m_{2}(m_1^2+m_3^2))\pa_{m_2}\nonumber\\&&
+(m_{1}^3+m_{2}^3-m_{3}(m_1^2+m_2^2))\pa_{m_3}]
\;J^{(g)}_{m_1m_2m_3}(x_1,x_2,x_3)\,,\label{7.64}\\
{B^{(3)}_2\;J^{(g)}_{m_1m_2m_3}(x_1,x_2,x_3)}&=&
 [(m_{2}^4+m_{3}^4-m_{1}^2(m_2^2+m_3^2))\pa_{m_1}
+(m_{1}^4+m_{3}^4-m_{2}^2(m_1^2+m_3^2))\pa_{m_2}\nonumber\\&&
+(m_{1}^4+m_{2}^4-m_{3}^2(m_1^2+m_2^2))\pa_{m_3}]
\;J^{(g)}_{m_1m_2m_3}(x_1,x_2,x_3)\,,\label{7.65}\\
s_0\;J^{(g)}_{m_1m_2m_3}(x_1,x_2,x_3)&=&
(\pa_{m_1}+\pa_{m_2}+\pa_{m_3})\;J^{(g)}_{m_1m_2m_3}(x_1,x_2,x_3)
\,,\label{7.66}\\
s_1\;J^{(g)}_{m_1m_2m_3}(x_1,x_2,x_3)&=&
(m_1\pa_{m_1}+m_2\pa_{m_2}+m_3\pa_{m_3})\;J^{(g)}_{m_1m_2m_3}(x_1,x_2,x_3)
\,,\label{7.67}\\
s_2\;J^{(g)}_{m_1m_2m_3}(x_1,x_2,x_3)&=&
(m_1^2\pa_{m_1}+m_2^2\pa_{m_2}+m_3^2\pa_{m_3})
\;J^{(g)}_{m_1m_2m_3}(x_1,x_2,x_3) \,.
\label{7.68}\eea
\end{theorem}
\Proof follows from the Proposition \ref{t4.5} (cf. formula (\ref{4.19}))
and the definition
of $A^{(k)}_{ij}$ (\ref{5.1}) (recall also that $B^{(k)}_{j}=$
$Res(A^{(k)}_{0j})$).
%
%
\section{Miscellaneous results}
\setcounter{equation}{0}
\noindent
Recall that
\beq
{B^{(k)}_j}=Res(S_0I_{j+k-1}-S_jI_{k-1})=
Res\left(\sum_{l<n}M^{(k)}_{ln}(D_n^j-D_l^j)\right)\,,
\label{8.1}\eeq
where
\beq
M^{(k)}_{ij}=x_iD_j^{k-1}-x_jD_i^{k-1}=-M^{(k)}_{ji}\,.
\label{8.2}\eeq
{}From the Proposition \ref{t4.1} one can derive the statement.
\begin{prop} \be [I_k,M^{(k)}_{ij}]=0\,. \ee \end{prop}
\vskip 1mm

\noindent
This means that we could repeat all the contruction described
in the previous Sections working with the operators $M^{(k)}_{ij}$
instead of the $S_i$ (and $A^{(k)}_{ij}$). The result is the same, i.e.
the operators $I_k$ commute not only to the $D_j$ but also to the additional
operators $M^{(k)}_{ij}$ being some sort of $(g,P)$-deformation
of the operators of `rotations'. The only difference between the operators
$A^{(k)}_{ij}$ and $M^{(k)}_{ij}$ is that the former ones commute with
permutations but the latter ones do not.

For instance for $k=2$ we have the following $(g,P)$-deformation
of the Euclidean e($N$) Lie algebra of generators $M_{ij}\equiv M^{(2)}_{ij}=
x_iD_j-x_jD_i$ and $D_k$:
\bea
{[M_{ij},P_{kl}]}&=&0\qquad \mbox{if}\qquad
\{i,j\}\cap\{k,l\}=\emptyset\,,\nonumber\\
{P_{ij}M_{ik}}&=&M_{jk}P_{ij}\,, \quad  \; k\neq j\,, \qquad
P_{ij}M_{ij}=M_{ji}P_{ij}\,,
\label{8.3}\eea
\pagebreak

\begin{eqnarray}
[M_{ij},M_{kl}]&=&
   \left(\delta_{ik}M_{lj}+\delta_{il}M_{jk}\right)
   \left(\id+g\sum_{m\neq i}P_{im}\right)
+\left(\delta_{jl}M_{ki}+\delta_{jk}M_{il}\right)
\left(\id+g\sum_{m\neq j}P_{jm}\right)\label{8.4}\\
&+&gM_{ij}P_{kl}\;\left(\delta_{ik}-\delta_{jl}+
\delta_{jk}-\delta_{il}\right)
-gM_{kl}P_{ij}\;\left(\delta_{ik}-\delta_{jl}-
\delta_{jk}+\delta_{il}\right)\,,\nonumber\\
{[M_{ij},D_j]}&=&D_i\left(\id+g\sum_{m\neq j}P_{jm}\right)+
gD_jP_{ij}\,.\label{8.5}
\end{eqnarray}
The relations (\ref{8.4})--(\ref{8.5}) should be supplemented
by the following additional relations for the case when
the indices in l.h.s.'s are mutually disjoint:
\begin{eqnarray}
[M_{ij},M_{kl}]&=&gM_{ik}P_{jl}+gM_{jl}P_{ik}
-gM_{il}P_{jk}-gM_{jk}P_{il}\,,\label{7}\\
{[M_{ij},D_{k}]}&=&gD_{j}P_{ik}-gD_{i}P_{jk}\,,\label{8.6}
\end{eqnarray}
i.e. if $\{i,j\}\cap\{k,l\}=
\emptyset$ and $\{i,j\}\cap\{k\}=\emptyset$, respectively.

For the case $g=0$ the relations (\ref{8.4}) and (\ref{8.5})
define the commutation relations for the generators
$M_{ij}$, $D_k$ of the Lie algebra
e$(N)$, i.e. the Lie algebra of the Lie group of mothions
of the $N$-dimensional Euclidean space. Keeping in mind this limit
we will treat the $PDM$-algebra as an one-parameter deformation of the Lie
algebra e$(N)$ (where $g$ is the parameter of deformation).
It is easy to see that the deformation involves
the permutation operators. Operator $\sum_{i}D_i^2$
 is a deformed Laplace operator
and, as well as in non-deformed case, is a Casimir operator of the
$PDM$-algebra. The very fact of the existence of the
$PDM$-algebra is an algebraic interpretation of the
super-integrability of the rational Calogero-Moser system,
i.e. there are not only $N-1$, but $2N-2$ independent additional
integrals of motion commuting with the Hamiltonian of
the rational Calogero-Moser problem.

{}From the Proposition \ref{t4.1} one can calculate that
the operators
\beq
J_+\equiv I_2=\sum_{i=1}^ND_i^2\,,\qquad J_3=\sum_{i=1}^N(x_iD_i+D_ix_i)\,,
\qquad J_-=\sum_{i=1}^Nx_i^2
\label{8.7}\eeq
have the sl(2) Lie algebra commutation relations and commute to the operators
$M_{ij}$ and permutations $P_{ij}$.
\begin{prop}\label{t8.2}
\beq
{[J_\pm,J_3]}=\pm 4J_\pm\,,\qquad [J_+,J_-]=2J_3\,,\qquad
{[J_{\pm,3},M_{ij}]}= [J_{\pm,3},P_{ij}]=0\,.\label{8.8}\eeq
\end{prop}
\vskip 1mm

\noindent
The Casimir operator of the sl(2) algebra (\ref{8.8}) has the form
\beq
C_2=\frac12 (J_+J_-+J_-J_+)-\frac14 J_3^2=\sum_{i<j}M_{ij}^2-
\left(gP+\frac{N}{2}\right)\left(gP+\frac{N}{2}-2\right)\,,
\label{8.9}\eeq
where the operator $P$ is the sum over all permutation operators:
\beq
P=\sum_{i<j}P_{ij}\,.
\label{8.10}\eeq
In order to get the purely differential operators from $M_{ij}$ and
$D_k$ one needs to consider the restriction of the
symmetric combinations of them like
\be
Res\left(\sum_{i,j}M_{ij}^{2r}\right)\qquad \mbox{or} \qquad
Res\left(\sum_{i,j}M_{ij}D_j^{r}\right)\,, \qquad r=1,2,3,\ldots\,.
\ee
Recall here that the additional operators $B^{(2)}_k$ look
like (cf. formula (\ref{8.1}))
\be
Res\left(\sum_{i,j}M_{ij}D_j^k\right)\,.
\ee
One could ask whether it is possible to find a {\it complete set
of commuting PDOs} only in terms of the symmetric combinations
of the operators $M_{ij}$ such that all these operators commute to
the Hamiltonian $H_2$. For $N=3$ it is possible.
\begin{theorem}
The following second order partial differential operators
are algebraically independent and mutually commuting:
\bea
{H_2}&=&Res(D_1^2+D_2^2+D_3^2)\label{8.11}\\
&=&\pa_1^2+\pa_2^2+\pa_3^2+\frac{2g}{x_{12}}\pa_{12}
+\frac{2g}{x_{13}}\pa_{13}+\frac{2g}{x_{23}}\pa_{23}
\,,\nonumber\\
{K_1}&=&Res(M_{12}^2+M_{23}^2+M_{31}^2)\label{8.12}\\
&=&{\cal M}_{12}^2+{\cal M}_{23}^2+{\cal M}_{31}^2
  +2g\left(\frac{x_1}{x_{23}}+\frac{x_2}{x_{31}}-\frac{x_1+x_2}{x_{12}}\right)
{\cal M}_{12}\nonumber\\
&&+2g\left(\frac{x_2}{x_{31}}+\frac{x_3}{x_{12}}-\frac{x_2+x_3}{x_{23}}\right)
{\cal M}_{23}
  +2g\left(\frac{x_3}{x_{12}}+\frac{x_1}{x_{23}}-\frac{x_3+x_1}{x_{31}}\right)
{\cal M}_{31}\,,\nonumber\\
{K_2}&=&Res(\{M_{12},M_{13}\}+\{M_{23},M_{21}\}+\{M_{31},M_{32}\})
\label{8.13}\\
&=&\{{\cal M}_{12},{\cal M}_{13}\}+\{{\cal M}_{23},{\cal M}_{21}\}
+\{{\cal M}_{31},{\cal M}_{32}\}
  +2g\frac{x_1-x_2-x_3}{x_{23}}(M_{13}-M_{12})\nonumber\\
&&+2g\frac{x_2-x_3-x_1}{x_{31}}(M_{21}-M_{23})
  +2g\frac{x_3-x_1-x_2}{x_{12}}(M_{32}-M_{31})\nonumber\\
&&-4g\frac{x_3}{x_{12}}M_{12}-4g\frac{x_1}{x_{23}}M_{23}
-4g\frac{x_2}{x_{31}}M_{31}\,,\nonumber\\
{{\cal M}_{ij}}&=&x_i\pa_j-x_j\pa_i\,.\nonumber\eea
Moreover, the spectral problem
\beq
H_2 \;\Psi =h_2\;\Psi\,,\qquad K_1 \;\Psi =k_1\;\Psi\,,\qquad
K_2 \;\Psi =k_2\;\Psi\,,\qquad h_2,k_1,k_2\in\RR
\label{8.14}\eeq
can be solved through a simple separation of variables (SoV)
which is some change of coordinates $x_i$'s.
\end{theorem}
\Proof The commutativity of the operators $H_2,K_{1,2}$ follows
from the Proposition \ref{8.2} and the SoV for $\Psi$ in (\ref{8.14})
is done by the same coordinate change by which Jacobi separated
variables in the corresponding Hamilton-Jacobi equation for
$H_2$.
\vskip 1mm

\noindent
The case $N=3$ is specific, there is obviously no such situation
for $N>3$. For general $N$ (and even for $N=3$) it is still open question
to make the SoV for the symmetric solution of the spectral problem
\beq
H_k\;\Psi_{\vec{m}}(\vec{x})=\sum_i m_i^k \;\Psi_{\vec{m}}(\vec{x})\,,
\qquad m_i\in\RR\,.\label{8.15}
\eeq
It is quite clear that such SoV will not be a simple change of coordinates
$x_i$'s but rather some integral transformation \cite{S}
\be
{\cal K}:\;\Psi_{\vec{m}}(\vec{x})\mapsto
\prod_{i=1}^N\;\psi_{\vec{m}}(y_i)\,.
\ee
See the work \cite{K} where the SoV was done for the trigonometric
generalisation of the 3-particle quantum Calogero-Moser system
(Calogero-Sutherland model) or, respectively, for the Jack polynomials
of the $A_2$ type. We hope that the results obtained there
and in the present work
will be helpful to work out the SoV for the spectral problem (\ref{8.15}).
%
%
\section*{Concluding remarks}
\noindent
Apart from the well-known and well-understood super-integrable systems
like $N$-dimensional harmonic oscillator and Coulomb problem, the rational
Calogero-Moser system leads to the non-linear algebra of hidden symmetry.

The construction of the `super-integrability table' showed in this paper
is quite general and can be applied to various generalisations such as
Calogero-Moser models associated to other classical root systems
(the model considered in this paper is associated to the $A_{N-1}$ root
system) and the Ruijsenaars-Schneider model which is a relativistic
generalisation. This work is in progress and will be published elsewhere.
\vskip 0.5cm

\noindent
{\it Acknowledgments} The author wishes to thank Ernie Kalnins and
Frank Nijhoff for valuable discussions. This work was partially supported
on the earlier stage through the research grant from the Waikato University,
Hamilton. The author acknowledges the hospitality of the Technical University
of Denmark.

\bibliographystyle{plain}

\end{document}